\documentclass[aps,prd,nofootinbib,twocolumn,floatfix,notitlepage]{revtex4-2}
\pdfoutput=1

\usepackage{graphicx}
\usepackage{psfrag,fancyhdr,epsfig}
\usepackage[breaklinks,colorlinks,urlcolor=blue,citecolor=blue,linkcolor=magenta]{hyperref}

\usepackage{amsmath}
\usepackage{xcolor}
\usepackage{tensor}

\definecolor{orange}{rgb}{0.9,0.2,0}
\definecolor{brown}{rgb}{0.7,0.3,0.2}
\definecolor{fuxia}{rgb}{1,0,1}
\definecolor{skyblue}{rgb}{0,0.1,0.9}
\definecolor{violetred}{rgb}{0.8,0.13,0.56}
\definecolor{deeppink}{rgb}{1.00,0.08,0.5}
\definecolor{pink}{rgb}{1.00,0.75,0.80}
\definecolor{orchid}{rgb}{0.85,0.44,0.84}
\definecolor{lightpink}{rgb}{1.00,0.71,0.76}
\definecolor{bluish}{rgb}{0,0.6,0.8}
\usepackage{slashed}
\usepackage{latexsym}
\usepackage{orcidlink}
\newcommand{\be}{\begin{equation}}
\newcommand{\ee}{\end{equation}}
\newcommand{\bear}{\begin{array}}
\newcommand{\eear}{\end{array}}

\definecolor{magmath}{rgb}{1,0,1} 

\def\a{\alpha}
\def\b{\beta}

\def\d{\delta}

\def\eps{\epsilon}
\def\g{\gamma}

\def\l{\lambda}
\def\m{\mu}
\def\n{\nu}

\def\r{\rho}
\def\s{\sigma}
\def\t{\tau}

\def\tns{\tensor}
\def\cR{\mathcal{R}}

\begin{document}

\title{Inflation in Weyl-invariant Einstein-Cartan gravity}

\author{Ioannis D. Gialamas\orcidlink{0000-0002-2957-5276}}
\email{ioannis.gialamas@kbfi.ee}
\affiliation{Laboratory of High Energy and Computational Physics, 
National Institute of Chemical Physics and Biophysics, R{\"a}vala pst.~10, Tallinn, 10143, Estonia}

\author{Kyriakos Tamvakis\orcidlink{0009-0007-7953-9816}}
\email{tamvakis@uoi.gr}
 \affiliation{Physics Department, University of Ioannina, 45110, Ioannina, Greece}

\begin{abstract}
We consider Weyl-invariant quadratic Einstein-Cartan gravity coupled to a scalar field and study the inflationary behavior of the coupled system of the scalar field and the pseudoscalar associated with the Holst invariant. We find that the model is characterized by effective single-field inflation occurring at small-field values and analyze its predictions which are in comfortable agreement with existing observations for a range of parameter values. 

\end{abstract}
\maketitle
 
%%%%%%%%%%%%%%%%%%%%%%%%%%%%%%%%%%%%%%%%%%%
\section{Introduction}
%%%%%%%%%%%%%%%%%%%%%%%%%%%%%%%%%%%%%%%%%%%
According to the cosmological inflation paradigm~\cite{Kazanas:1980tx,Sato:1980yn,Guth:1980zm,Linde:1981mu} the Universe has undergone a phase of quasi-de Sitter expansion during which the quantum fluctuations of gravitational and matter fields have been upgraded to the cosmological perturbations~\cite{Starobinsky:1979ty,Mukhanov:1981xt,Hawking:1982cz,Starobinsky:1982ee,Guth:1982ec,Bardeen:1983qw} that are responsible for the large-scale structure of the Universe. While the detailed mechanism of inflation is not known, the standard way to model it is in terms of a scalar field that provides the necessary vacuum energy and, through its quantum fluctuations, provides the seed for the spatial inhomogeneities presently detected in the the cosmic microwave background. The  theoretical framework of cosmological considerations is general relativity (GR) as a theory of gravity and quantum field theory as a theory of matter interactions. Although gravity, being weak well below the Planck energy scale, is treated classically within GR, the gravitating matter interactions should be treated in full quantum fashion. As a result, the effective classical theory of gravity includes corrections in the form of nonminimal couplings of scalar fields to curvature or higher-power curvature terms. Alternative formulations of gravity like the {\textit{metric-affine}} formulation~\cite{Hehl:1994ue,Baldazzi:2021kaf,Aoki:2023sum} in which the connection is an independent variable in addition to the metric, although they coincide with the standard (metric) formulation in the framework of the standard Einstein-Hilbert action of GR, they lead to different predictions when the above curvature corrections~\cite{Sotiriou:2006qn} or couplings to matter~\cite{Rigouzzo:2022yan,Rigouzzo:2023sbb} are included. This, among other motivations, has prompted an interest in the study of alternative formulations of gravity and especially their inflationary behavior, since inflation provides a direct link between predictions and observations.

A case where the metric-affine formulation of gravity differs from the standard (metric) one is that of the presence of quadratic corrections of the curvature in the action. In particular, quadratic terms of the Ricci scalar and the Holst invariant~\cite{Holst:1995pc} lead to an equivalent metric theory that includes an extra pseudoscalar\footnote{As we will discuss later, the Holst invariant is formed by contracting the Riemann tensor with the Levi-Civita pseudotensor. Consequently, the scalar field associated with the Holst invariant is a pseudoscalar, i.e. it changes sign under a parity inversion.} degree of freedom,  with numerous applications in inflation, gravitational waves~\cite{Battista:2021rlh,Elizalde:2022vvc,Battista:2022hmv,DeFalco:2023tcp,DeFalco:2024ojf}, and dark matter phenomenology~\cite{BeltranJimenez:2019hrm,Langvik:2020nrs,Pradisi:2022nmh,Gialamas:2022xtt,Gialamas:2023emn,DiMarco:2023ncs,Inagaki:2024ltt,Racioppi:2024zva}. This model is equivalent to the special case of the so-called \textit{Einstein-Cartan} (EC) formulation in which {\textit{torsion}} has an important role~\cite{Shaposhnikov:2020gts,Shaposhnikov:2020aen,Karananas:2021zkl,Karananas:2021gco,Piani:2022gon,Piani:2023aof,He:2024wqv},  while nonmetricity can be set to zero due to an extended projective symmetry~\cite{Afonso:2017bxr,Iosifidis:2018jwu,Barker:2024dhb} of the curvature scalars. On the other hand, since quadratic curvature terms are central in this model, their property of being invariant under Weyl transformations naturally leads one to consider a Weyl-invariant version of the model~\cite{Edery:2014nha,Edery:2015wha,Ghilencea:2018dqd,Ghilencea:2020piz,Ghilencea:2021jjl,Ghilencea:2021lpa,Shtanov:2023lci,Wang:2022ojc,Yang:2022icz,Burikham:2023bil,Condeescu:2023izl,Harko:2024fnt,Condeescu:2024cjh,Ghilencea:2024usf}. Ample motivation for
doing this is provided by the fact that Weyl invariance implies the absence of any mass scales. Classical scale invariance, necessarily broken at the quantum level, has been thought to be related to the observed vastly different phenomenological energy scales in physics~\cite{Shaposhnikov:2008xb,Shaposhnikov:2008xi,Garcia-Bellido:2011kqb,Blas:2011ac,Bezrukov:2012hx,Khoze:2013uia,Csaki:2014bua,Kannike:2014mia,Kannike:2015apa,Rinaldi:2015uvu,Ferreira:2016vsc,Marzola:2016xgb,Karananas:2016kyt,Kannike:2016wuy,Ferreira:2016wem,Ghilencea:2016dsl,Rubio:2017gty,Benisty:2018fja,Ferreira:2018qss,Karam:2018mft,Ferreira:2018itt,Ferreira:2018qss,Gialamas:2020snr,Shaposhnikov:2018jag,Iosifidis:2018zwo,Casas:2018fum,Gialamas:2021enw,Cecchini:2024xoq}. Weyl-invariant Einstein-Cartan gravity is directly related to the {\textit{Weyl geometrical formulation}} featuring a torsion vector gauge field~\cite{Barker:2024ydb,Barker:2024goa}. 

In the present article, we consider Weyl-invariant EC gravity nonminimally coupled to a fundamental scalar with a quartic potential and study its inflationary properties (see also~\cite{Ghilencea:2018thl,Tang:2019olx,Ghilencea:2019rqj,Ghilencea:2020rxc,Hu:2023yjn}) in the presence of the axionlike pseudoscalar field related to the Holst invariant.  Recently, it has been suggested that this model, for a certain range of its parameters, could provide a possible solution to the strong \textit{CP} problem of particle physics~\cite{Karananas:2024xja}.

We consider the coupled system of the two arising scalar fields in a Friedmann-Robertson-Walker (FRW) background and show that the system evolves rapidly along a specific trajectory in the two-field space into an effective single-field system for which we proceed to study slow-roll inflation. The ascertained inflationary behavior differs substantially from the standard Higgs inflation, being associated with the small-field region of the inflaton (i.e., inflation occurs at field values smaller than the vacuum expectation value of the field). This behavior should also be contrasted to what happens in the case of (Weyl-noninvariant) quadratic metric-affine gravity where the inflationary plateau arises at the large-field region~\cite{Enckell:2018hmo,Antoniadis:2018ywb,Tenkanen:2019jiq,Edery:2019txq,Gialamas:2019nly,Pradisi:2022nmh,Salvio:2022suk,Gialamas:2022xtt}. In addition the pseudoscalar associated with the Holst invariant leads to significant deviations when compared to the results in~\cite{Ghilencea:2019rqj}. The inflationary predictions of the model for the spectral index and the tensor-to-scalar ratio turn out to be in comfortable agreement with observations for a significant range of the parameters. 

The article is organized as follows:
In Sec.~\ref{sec:framework}, we describe the theoretical framework of Weyl-invariant EC gravity. In Sec.~\ref{sec:themodel}, we consider the Weyl-invariant coupling to a scalar field analyzing the equivalent metric form of the theory that includes the additional pseudoscalar that arises due to the presence of the Holst invariant in the gravitational action. In Sec.~\ref{sec:inflation}, we proceed to study the inflationary properties of the scalar sector which is dynamically reduced to an effective single-field system and analyze its observable predictions in the light of present data. In Sec.~\ref{sec:conclusions}, we state briefly our conclusions.

\section{Framework}
\label{sec:framework}

The EC formulation of gravity consists in promoting the Poincar{\'e} group into a local symmetry. In this framework translations and Lorentz transformations correspond to the curvature and torsion gauge fields, respectively. Their field strengths can be  represented by the {\textit{affine curvature}} and the {\textit{affine torsion}} 
\begin{align}
\tns{\cR}{^\r_\s_\m_\n}=&\,\, \partial_{ \mu}\tns{\Gamma}{^\r_\n_\s}-\partial_{ \nu}\tns{\Gamma}{^\r_\m_\s}+\tns{\Gamma}{^\r_\m_\l}\tns{\Gamma}{^\l_\n_\s}-\tns{\Gamma}{^\r_\n_\l}\tns{\Gamma}{^\l_\m_\s}\,, \nonumber
\\
\tns{T}{^\r_\m_\n}=&\,\, \tns{\Gamma}{^\r_\m_\n}-\tns{\Gamma}{^\r_\n_\m}\,,
\end{align}
expressed in terms of \textit{affine connection} $\Gamma^{ \rho}_{\,\,\mu\nu}$.
The torsion can be decomposed as\footnote{The Levi-Civita tensor is given by $\eps_{\a\b\g\d} =\sqrt{-g} \mathring{\varepsilon}_{\a\b\g\d}$, where $ \mathring{\varepsilon}_{\a\b\g\d}$ is the Levi-Civita symbol with $ \mathring{\varepsilon}_{0123}=1$.}
\be 
T_{\mu\nu\rho}=\frac{1}{3}\left(g_{ \mu\rho}T_{ \nu}-g_{ \mu\nu}T_{\rho}\right)+\frac{1}{6}\epsilon_{ \mu\nu\rho\sigma}\hat{T}^{ \sigma}+\tau_{ \mu\nu\rho}\,,
\ee
in terms of a \textit{torsion vector} $T_{ \mu}=\tns{T}{^\r_\m_\r}$, an {\textit{axial vector}} $\hat{T}^{ \mu}=\epsilon^{ \mu\nu\rho\sigma}T_{ \nu\rho\sigma}$ part
and a \textit{tensorial part} $\tau_{ \mu\nu\rho}$ (with $\tns{\t}{^\a_\m_\a}=\tns{\t}{^\a_\a_\m}=\epsilon^{ \mu\nu\rho\sigma}\tau_{ \nu\rho\sigma}=0$).

Weyl transformations consist of local metric rescalings $g_{ \mu\nu}\rightarrow\Omega^{-2}g_{ \mu\nu}$, while
Weyl invariance, corresponding to the absence of mass scales at the classical level and necessarily broken at the quantum level~\cite{Shaposhnikov:2008xi,Ghilencea:2016dsl,Ferreira:2018itt}, is a symmetry that has often been invoked in explaining the vast difference of arising mass scales in physics. Under Weyl rescalings the curvature ${\cal{R}}^{ \rho}_{\,\,\sigma\mu\nu}$ is invariant, while the torsion transforms as\footnote{The torsion vector and tensor parts transform as
$$ T_{\mu}\rightarrow\,T_{ \mu}-3\frac{\partial_{ \mu}\Omega}{\Omega},\,\,\,\,\,\,\hat{T}_{ \mu}\rightarrow\,\hat{T}_{ \mu},\,\,\,\,\,\,\tau_{ \mu\nu\rho}\rightarrow\,\Omega^{-2}\tau_{ \mu\nu\rho}\,.$$}
\be 
\tns{T}{^\r_\m_\n}\rightarrow\tns{T}{^\r_\m_\n} - 
 \frac{\partial_{ \mu}\Omega}{\Omega}\delta^{ \rho}_{ \nu}\,+\,\frac{\partial_{ \nu}\Omega}{\Omega}\delta^{ \rho}_{\,\mu}\,.
\ee

A theory of gravity in the EC framework has to be formulated in terms of an action that contains at most quadratic terms of the curvature and torsion in order to avoid higher derivatives. Furthermore, Weyl invariance, forbidding dimensionful coefficients, disallows linear terms. Thus, only quadratic terms of the curvature are allowed. General quadratic curvature terms are known to introduce new unwanted degrees of freedom such as higher-spin ghosts~\cite{Stelle:1976gc}. Nevertheless, these can be avoided in the special case of a quadratic term of the scalar curvature. In fact in this framework there are two such curvature scalars, namely the {\textit{Ricci scalar}} ${\cal{R}}$ and the {\textit{Holst invariant}} $\widetilde{\cR}$, defined as
\be 
{\cal{R}}= \tns{\cR}{^\m^\n_\m_\n}\,,\qquad \widetilde{\cR}=\epsilon^{\mu\nu\rho\sigma}\cal{R}_{ \mu\nu\rho\sigma}\,.
\ee 
Therefore, a Weyl-invariant EC pure gravity action can contain only ${\cal{R}}^2$, $\widetilde{\cR}^2$ and possibly ${\cal{R}}\widetilde{\cR}$ (see the Appendix~\ref{appendix}). 

Keeping things simple, we start with a Weyl-invariant Einstein-Cartan gravitational action 
\be 
\mathcal{S}_0=\int{\rm d}^4x\sqrt{-g}\left\{\frac{\gamma}{4}{\cal{R}}^2+\frac{\delta}{4}\widetilde{\cR}^2\right\}\,,
\label{SO}
\ee
where $\g$ and $\d$ are constant parameters with dimensions of mass$^{-2}$.
This can be rewritten in terms of two auxiliary fields\footnote{The Weyl transformations of the auxiliary fields are $\chi,\,\zeta\rightarrow\,\Omega^2\chi,$ and $\Omega^2\zeta\,.$} $\chi$ and $\zeta$ as (see~\cite{Gialamas:2022xtt} for more details)
\be 
\hspace{-0.15cm}\mathcal{S}_0=\int{\rm d}^4x\sqrt{-g}\left\{\frac{\gamma}{2}\chi{\cal{R}}+\frac{\delta}{2}\zeta\widetilde{\cR}-\left(\frac{\gamma}{4}\chi^2+\frac{\delta}{4}\zeta^2\right)\right\}\,.
\ee 

\section{The model}
\label{sec:themodel}

Next, we consider coupling EC gravity to a fundamental scalar in a Weyl-invariant manner, adding to $\mathcal{S}_0$ the matter action
\begin{align}
\Delta\mathcal{S}=&\int{\rm d}^4x\sqrt{-g}\bigg\{-\frac{1}{2}(\tilde{D}_\m h)^2-\frac{\lambda}{4}h^4+C\hat{T}^{ \mu}\hat{T}_{ \mu}h^2\nonumber
\\
&+\frac{\xi}{2}h^2{\cal{R}}+\frac{\widetilde{\xi}}{2}h^2\widetilde{\cR}\bigg\}\,,
\label{DELTAS}
\end{align}
where $h$ transforms under Weyl transformations as $h\,\rightarrow\,\Omega h\,$. Note that the fundamental scalar $h$ is coupled to EC gravity through terms involving the Ricci scalar, the Holst invariant, and the torsion and axial vectors.
The first term in~\eqref{DELTAS} features the so-called \textit{Weyl-covariant derivative}~\cite{Karananas:2021gco}
\be \tilde{D}_{ \mu}h=\partial_{ \mu}h\,+\,\frac{1}{3}T_{ \mu}h\,,\ee
transforming as $\tilde{D}_\m h\rightarrow \Omega\tilde{D}_\m h$, while the third term is simply allowed by Weyl invariance and is scaled by an arbitrary constant $C$. The parameters $\xi$ and $\widetilde{\xi}$ parametrize the nonminimal couplings of $h$ to the curvature scalars and are typically included in the action because they are also Weyl invariant and have dimensions of mass$^4$.

The potential is also Weyl invariant if we assume it to be quartic, i.e. $ \l h^4/4$. Since $\t^{\m\n\r}$ appears quadratically in both $\cR$ and $\widetilde{\cR}$, its equations of motion will have $\t^{\m\n\r}=0$ as a solution. Therefore, terms of the form $\sim h^2 \t_{\m\n\r}\t^{\m\n\r}$ and $\sim h^2 e^{\m\n\r\s} \t_{\l\m\n} \tns{\t}{^\l_\r_\s}$, while potentially relevant, give no contribution on shell.

It is possible to obtain a metric equivalent description of the action $\mathcal{S}=\mathcal{S}_0+\Delta\mathcal{S}$ making use of the expressions
\begin{align}
{\cal{R}}=&\,\,R[g]+2D_{ \mu}T^{ \mu}-\frac{2}{3}T_{ \mu}T^{ \mu}+\frac{1}{24}\hat{T}^{\mu}\hat{T}_{\mu}+\frac12\tau_{ \mu\nu\rho}\tau^{ \mu\nu\rho}\,, \nonumber
\\
\widetilde{\cR}=&\,\,-D_{ \mu}\hat{T}^{ \mu}+\frac{2}{3}\hat{T}_{ \mu}T^{ \mu}+\frac{1}{2}\epsilon^{ \mu\nu\rho\sigma}\tau_{ \lambda\mu\nu}\tns{\t}{^\l_\r_\s}\,,
\end{align}
where $R[g]$ and $D_{ \mu}$ are constructed in terms of the metric (Levi-Civita) connection.

Weyl invariance allows for a {\textit{gauge choice}} $\chi=M_P^2/\g$~\cite{Karananas:2024xja} leading to the following form of the action (in Planck mass units)\footnote{When including fermions, the Weyl ivariant coupling of $\hat{T}^\m$ to the axial fermionic current, $\Bar{\Psi}\g_5\g_\mu\Psi$, results in a coupling of the pseudoscalar $\zeta$ to the axial anomaly that leads to a possible solution to the strong $CP$ problem albeit for extreme parameter values $y_\zeta\sqrt{\delta}\sim 10^{43}$~\cite{Karananas:2024xja}. Here, $y_\zeta$ denotes the coupling constant for the interaction between the axial torsion vector and the axial current.}
\begin{align}
\mathcal{S} &=\int{\rm d}^4x\sqrt{-g} \bigg\{ \frac{1}{2}(1+\xi h^2)\bigg(R[g]-\frac{2}{3}T^2 +\frac12\tau_{ \mu\nu\rho}\tau^{ \mu\nu\rho} \nonumber
\\
& +\frac{1}{24}\hat{T}^{\mu}\hat{T}_{\mu}\bigg)+\frac{1}{2}(\delta\zeta+\widetilde{\xi}h^2)\left(\frac{2}{3}\hat{T}^\m T_\m+\frac{\epsilon^{ \mu\nu\rho\sigma}}{2}\tau_{ \lambda\mu\nu}\tns{\t}{^\l_\r_\s}\right) \nonumber
\\
& +\frac{1}{2}\hat{T}_\m\partial^\m(\delta\zeta+\widetilde{\xi}h^2) -\frac{1}{2}(\partial_\m h)^2-\left(\frac{1}{3}+
 2\xi\right) h T_\m\partial^\m h \nonumber
\\
&-\frac{1}{18}h^2T_\m T^\m+Ch^2\hat{T}_\m \hat{T}^\m -\frac{1}{4\gamma}-\frac{\delta}{4}\zeta^2-\frac{\lambda}{4}h^4\bigg\}\,. {\label{S}}
\end{align}
Since no derivatives of $T,\hat{T},$ or $\tau$ appear in~\eqref{S}, these variables do not represent any propagating degrees of freedom, their equations of motion being purely algebraic. In particular, the $\tau$ equation of motion yields simply $\tau_{ \mu\nu\rho}=0$. Varying with respect to $T$ and $\hat{T}$ we obtain a linear system, the solution of which is substituted back into the action. Rescaling to the Einstein frame through $g_{ \mu\nu}\rightarrow\left(1+\xi h^2\right)^{-1}g_{ \mu\nu}$, after quite a bit of calculation, we obtain the action in a diagonal form in terms of a canonical fundamental scalar $H$ as\footnote{The variables $H$ and $\Phi$ are given by in terms of the original scalar variables $h$ and $\zeta$ by~\cite{Karananas:2024xja} $\Phi= (1+6\xi)^{-1}\ln[(6+(1+6\xi)h^2)/(3\widetilde{\xi}-\d\zeta(1+6\xi)/2)]$ and $H=\sqrt{6} \tanh^{-1}\left[h/\sqrt{6+(1+6\xi)h^2}\right]$. Note that $H\rightarrow 0$ for $h\rightarrow 0$, while for large values of $\xi h^2\gg 1$ the canonical field reaches a maximum $H\sim \sqrt{6}\tanh^{-1}(1/\sqrt{1+6\xi})$.}
\be
 \mathcal{S}=\frac12\int{\rm d}^4x\sqrt{-g}\left\{R[g]-(\partial_\m H)^2-K_\Phi(\partial_\m\Phi)^2 -2V\right\}\,,
 \label{eq:S_2field}
\ee
where
\begin{align}
K_\Phi=& \, \frac{3}{2}\cosh^4(H/\sqrt{6})\Bigg[ \left(\frac{1-\frac{\widetilde{\xi}}{2}e^{(1+6\xi)\Phi}}{(1+6\xi)}\right)^2 \cosh^2(H/\sqrt{6}) \nonumber
\\
& +\frac{e^{2(1+6\xi)\Phi}}{(24)^2}\left(1+144C\sinh^2(H/\sqrt{6})\,\right)\Bigg]^{-1}\,,
\\[0.2cm]
V = &\, 9\lambda\sinh^4(H/\sqrt{6})+\frac{1}{4\gamma}\left(1-6\xi\sinh^2(H/\sqrt{6})\right)^2 \nonumber
\\
 & + \frac{9}{\delta}\frac{e^{-2(1+6\xi)\Phi}}{(1+6\xi)^2}\Bigg[2\cosh^2(H/\sqrt{6}) \nonumber
\\
& -\widetilde{\xi}e^{(1+6\xi)\Phi}\left(1-6\xi\sinh^2(H/\sqrt{6})\right)\Bigg]^2\,.
\label{eq:pot_V}
\end{align}
The potential $V(H,\Phi)$ has a {\textit{minimum line}} in the $H/\Phi$ two-field space along
\be 
e^{(1+6\xi)\Phi}=\frac{2}{\widetilde{\xi}}\frac{\cosh^2(H/\sqrt{6})}{[1-6\xi\sinh^2(H/\sqrt{6})]}\,.
\label{eq:min_phi}
\ee 
Along the minimum line~\eqref{eq:min_phi} the contribution of $\Phi$ to the potential is removed but it still has a contribution to the kinetic terms. Substituting~\eqref{eq:min_phi} back into the action~\eqref{eq:S_2field}, we obtain the single-field action
\be
\hspace{-0.18cm} \mathcal{S}=\int{\rm d}^4x\sqrt{-g}\left\{\frac{R[g]}{2}-\frac{\widetilde{K}(H)}{2}(\partial_\m H)^2-U(H)\right\}\,,
 \label{eq:S_1field}
\ee
where
\begin{align}
\label{eq:Ksingle}
\widetilde{K}(H) = & \,1+ \sinh^2(H/\sqrt{6})\bigg[\sinh^4(H/\sqrt{6}) 
\\
&\hspace{-1.4cm}+(1/12\widetilde{\xi})^2 \cosh^2(H/\sqrt{6})\left(1+144C \sinh^2(H/\sqrt{6})\right) \bigg]^{-1}\,, \nonumber
\\[0.2cm]
U(H) = & \,9\lambda\sinh^4(H/\sqrt{6})+\frac{1}{4\gamma}\left(1-6\xi\sinh^2(H/\sqrt{6})\right)^2\,.
\label{eq:Potsingle}
\end{align}
The resulting single-field potential exhibits a plateau region near the origin ($H\sim 0$) and an absolute minimum at larger field values, followed by exponential growth.
Note that the potential, expressed in terms of the original field $h$ has a form analogous to the one encountered in standard Higgs inflation~\cite{Bezrukov:2007ep}, namely
\be
U(h)=\frac{\lambda}{4}\frac{h^4}{(1+\xi h^2)^2}+\frac{1}{4\gamma(1+\xi h^2)^2}\,.
\ee
Nevertheless, the kinetic term, expressed in terms of $h$, for large field values, $\xi h^2\gg 1$, behaves as $\widetilde{K}(\nabla H)^2\sim h^{-6}(\nabla h)^2$, in contrast to standard Higgs inflation~\cite{Bezrukov:2007ep}. As a result, any inflationary behavior should be associated with the \textit{small $h\sim H$ region}.

\begin{figure}[t!]
\centering
\includegraphics[width=0.48\textwidth]{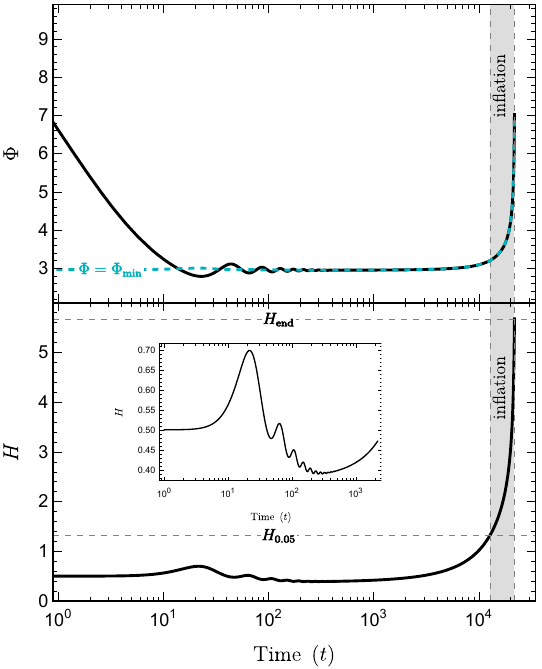}
\caption{The evolution of the scalar fields $\Phi$ (upper) and $H$ (lower). The teal dashed line in the upper plot corresponds to the minimum line~\eqref{eq:min_phi}. The inflationary region is highlighted in gray, with $H_{\rm end}$ and $H_{0.05}$ representing the field values at the end of inflation and at the pivot scale $k_\star = 0.05\, {\rm Mpc}^{-1}$, respectively. The parameters used are $\xi=0.004\,,\, \widetilde{\xi}=0.1\,,\, C=0\,,\, \d=4\times 10^6$, and $\l\g = 10^{-10}/36$. } 
\label{fig:1}
\end{figure}
 
\section{Inflation}
\label{sec:inflation}

In this section, we examine the inflationary behavior of the model, with the goal to constrain its parameter space by comparing its inflationary predictions to the latest observational bounds~\cite{Planck:2018jri,BICEP:2021xfz}.

Assuming an FRW background (with $\mathcal{H}$ being the Hubble constant) the equations of motion arising from the original two-field action~\eqref{eq:S_2field} read
\begin{align}
K_\Phi\ddot{\Phi}+3{\cal{H}}K_\Phi\dot{\Phi}+\dot{H}\dot{\Phi}\frac{\partial K_\Phi}{\partial H}+\frac{\dot{\Phi}^2}{2}\frac{\partial K_\Phi}{\partial\Phi} &=-\frac{\partial V}{\partial\Phi}\,,\nonumber
\\
\ddot{H}+3{\cal{H}}\dot{H}-\frac{\dot{\Phi}^2}{2}\frac{\partial K_\Phi}{\partial H} &=-\frac{\partial V}{\partial H}\,.
\label{eq:frw_eoms}
\end{align}
We proceed to solve numerically the system of equations~\eqref{eq:frw_eoms}. In Fig.~\ref{fig:1} we show the temporal evolution of the scalar fields $\Phi$ and $H$ arising from the solution of the above system. It is clear that by the time $H$ reaches the inflationary plateau, $\Phi$ has already fallen into the valley defined by the minimum line~\eqref{eq:min_phi} (teal dashed line), following a period during which both fields were oscillating around their minima (for the parameter values used in Fig.~\ref{fig:1}, the minima occur at $H_{\rm min} \simeq 6.3$ and $\Phi(H_{\rm min})\simeq 16.3$; however, these values are not displayed in the plot for clarity). This observation supports the conclusion that the model effectively reduces to a single-field inflationary scenario. In Fig.~\ref{fig:1}, the parameters are chosen to be $\xi=0.004\,,\, \widetilde{\xi}=0.1\,,\, C=0\,,\, \d=4\times 10^6$, and $\l\g = 10^{-10}/36$. The specific value of the product $\l\g$ has been selected to match that used in~\cite{Ghilencea:2019rqj}. Finally, our numerical analysis shows that $\Phi$ rolls towards the valley more quickly as the parameters $\widetilde{\xi}$ and $\d$ decrease. As discussed earlier, a possible solution to the strong \textit{CP} problem~\cite{Karananas:2024xja} is achieved for very large values of the product $y_\zeta\sqrt{\d}$. Consequently, a typical value of the axial torsion vector-axial current coupling $y_\zeta$ would require an exceptionally large $\d$. Such large values might slow down the rolling of the pseudoscalar $\Phi$ toward its minimum direction. However, this does not necessarily imply that the successful single-field inflation predictions, which we will present later, are compromised. To draw definitive conclusions, a two-field analysis is required.

Therefore, for the inflationary period it would be sufficient to study the single field described by the action along the minimum direction~\eqref{eq:min_phi} given by~\eqref{eq:S_1field}. Inflation on the same single-field model, albeit with $\widetilde{K}(H) = 1$, was analyzed in~\cite{Ghilencea:2019rqj} (see also~\cite{Ferreira:2019zzx, Ghilencea:2020rxc}). However, in our case, the deviation of the kinetic function $\widetilde{K}(H)$ from unity can lead to significantly different inflationary predictions.

\subsubsection{Observables}
Regarding the cosmological observables constrained by recent observations~\cite{Planck:2018jri,BICEP:2021xfz}, we start with the amplitude of the primordial curvature perturbation on super-Hubble scales produced by single-field slow-roll inflation. This amplitude is expressed as $A_s^\star = \mathcal{H}_\star^2/(8\pi^2\eps_1^\star)$ and has been constrained to the value $A_s^\star \simeq 2.1\times 10^{-9}$ at the pivot scale $k_\star = 0.05\, {\rm Mpc}^{-1}$~\cite{Planck:2018jri}.  The tensor-to-scalar ratio $(r)$ and the spectral index of the scalar power spectrum $(n_s)$ are given by
$r= 16\eps_1 $ and $ n_s = 1-2\eps_1-\eps_2\,,$ respectively.
In these expressions, we have used the Hubble flow functions, also known as slow-roll parameters, defined as $\eps_1 = -{\rm d}\ln\mathcal{H}/{\rm d} N$ and $\eps_2 = {\rm d}\ln\eps_1/{\rm d} N$, where ${\rm d}N =\mathcal{H}{\rm d}t$. It is important to note that due to the complexity of the kinetic function~\eqref{eq:Ksingle}, providing analytic approximations for the observables is quite challenging. Therefore, we solve the equations of motion numerically and employ the Hubble flow functions in our computations of the observables, rather than relying on the traditional potential slow-roll parameters as is common in similar analyses.

During inflation, the evolution of the scalar field is closely related to the number of $e$-folds, denoted as $N_\star$. Assuming instantaneous thermalization of the Universe immediately after inflation, $N_\star$ is given by~\cite{Liddle:2003as}
\be
\label{eq:efolds}
N_\star = 66.5 - \ln \left(\frac{k_\star}{a_0 \mathcal{H}_0}\right) +\frac14\ln\left(\frac{9\mathcal{H}_\star^4}{\rho_{\rm end}}\right)\,,
\ee
where $a_0 $ and $\mathcal{H}_0$ are the present-day values of the scale factor and the Hubble rate, respectively, and $\rho_{\rm end}$ is the energy density of the inflaton field at the end of inflation which is equal to $\rho_{\rm end} = 3U(H_{\rm end})/2$.

\begin{figure}[t!]
\centering
\includegraphics[width=0.48\textwidth]{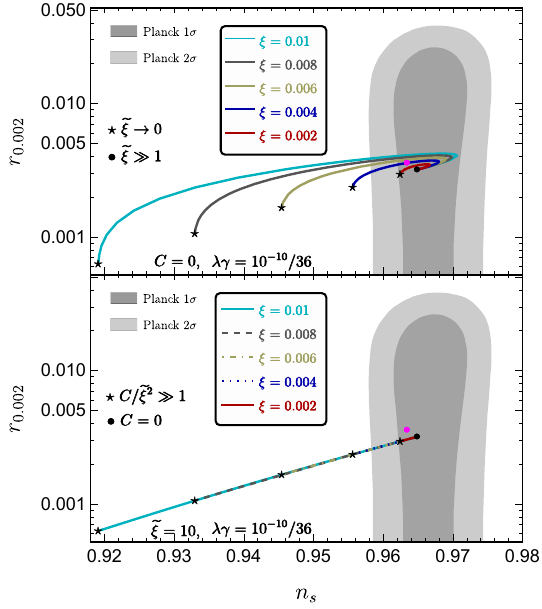}
\caption{Model predictions using pivot scales $k_\star = 0.05\, {\rm Mpc}^{-1}$
for the spectral index and  $k_\star = 0.002\, {\rm Mpc}^{-1}$
for the tensor-to-scalar ratio. Shaded gray regions indicate the $68\%$ and $95\%$ confidence intervals, based on the latest data from Planck, BICEP/Keck and BAO~\cite{Planck:2018jri,BICEP:2021xfz}. The number of $e$-folds is constrained by Eq.~\eqref{eq:efolds} to be $N_\star \simeq 55.3-55.8$ across all curves. The magenta bullet ($\color{magmath} \bullet$) represents the predictions derived from the sample parameters used in Fig.~\ref{fig:1}.
}
\label{fig:2}
\end{figure}
\subsubsection{Model analysis}
The single-field potential described by Eq.~\eqref{eq:Potsingle} features a symmetric plateau region on both sides of the origin, $H=0$, with a minimum occurring at larger field values~\cite{Ghilencea:2019rqj}. The potential values at the origin and at the minimum are given by
\be
U(0) = \frac{1}{4\gamma}\,, \qquad \text{and} \qquad U(H_{\rm min}) = \frac{1}{4\g+4\xi^2/\l}\,,
\ee
with $H_{\rm min} = \sqrt{6} \sinh^{-1}\left(\sqrt{\xi/(6\g\l+6\xi^2)}\right)$. For successful inflation, the initial potential energy must be significantly greater than the energy at the minimum, i.e., $U(0) \gg U(H_{\rm min}) \simeq 0$. This ensures that inflation can terminate, allowing the standard reheating epoch to occur. This requirement constrains the model parameters to the region $\l\g \ll \xi^2$. Within this parameter space, the minimum of the potential is approximately at $H_{\rm min} \sim \sinh^{-1}(1/\sqrt{\xi})$. Since the function $\sinh^{-1}$ is monotonically increasing, a large value of $\xi$ shifts $H_{\rm min}$ closer to the origin, thereby reducing the plateau region of the potential. Additionally, small values of $\xi$ help address the issue of violating perturbative unitarity at the scale $M_P/\xi$ ($M_P/\sqrt{\xi}$) in standard~\cite{Barbon:2009ya} (metric-affine~\cite{Bauer:2010jg}) Higgs inflation. Thus, large values of $\xi$ are generally disfavored.

Assuming that inflation begins near the origin, we can estimate the value of the parameter $\g$. Using the relation $H_\star^2 \simeq U(0)/3$ and the constraint $r_{0.05}<0.036$~\cite{BICEP:2021xfz}, together with $A_s^\star \simeq 2.1\times 10^{-9}$, we find that $\g \gtrsim 2.24\times 10^8$. Therefore, large values of $\g$ are compatible with the condition $\l\g \ll \xi^2$ only if the quartic coupling $\lambda$ is extremely small.

As discussed earlier, the kinetic function~\eqref{eq:Ksingle} plays a key role in the significant deviations observed in our model when compared to the results in~\cite{Ghilencea:2019rqj}. Nevertheless, the predictions from~\cite{Ghilencea:2019rqj} can be recovered in specific regions of the parameter space. This agreement becomes evident in the limit $\widetilde{\xi} \rightarrow 0$, where $\widetilde{K}(H) \simeq 1  + \mathcal{O}(\widetilde{\xi}^2)$, leading to inflationary predictions in our model that align with those of~\cite{Ghilencea:2019rqj}. Furthermore, a similar conclusion holds when $\widetilde{\xi}$ is relatively large ($\widetilde{\xi}\gtrsim 10$), provided that $C\gg \widetilde{\xi}^2$. In this regime, we find that
\be
\widetilde{K}(H) \simeq 1 +\frac{\widetilde{\xi}^2/C}{\cosh^2(H/\sqrt{6})} \simeq 1\,,
\label{eq:KClimit}
\ee
since $1/\cosh^2(H/\sqrt{6})$ is bounded from above by $1$.

Alongside the previously mentioned bound on $r$, the scalar spectral index is constrained to be $n_s = 0.9649 \pm 0.0042$~\cite{Planck:2018jri}.
In Fig.~\ref{fig:2}, we present the model's predictions using pivot scales $k_\star = 0.05\, {\rm Mpc}^{-1}$ for the spectral index and $k_\star = 0.002\, {\rm Mpc}^{-1}$ for the tensor-to-scalar ratio. The number of $e$-folds is constrained by Eq.~\eqref{eq:efolds} to be $N_\star \simeq 55.3-55.8$ across all curves.

In the upper panel, we analyze the case $C=0$. The stars ($\star$) indicate the predictions from~\cite{Ghilencea:2019rqj} where $\widetilde{\xi}=0$. As $\widetilde{\xi}$ increases, the predictions shift to the right and align with the observational bounds represented by the gray shaded regions. 
Our numerical analysis shows the existence of a  fixed point arising for $\widetilde{\xi}$ sufficiently large. All curves  are frozen to a particular point (denoted by $\bullet$) regardless of the value of the parameter $\xi$, and remain unchanged with any further increase in $\widetilde{\xi}$. Although we have not been able to derive an analytic derivation for the fixed point in the $n_s-r$ plane, its existence is supported by the relatively weak $\xi$ dependence of the potential in the small-field region where inflation takes place. At this point, the predictions are $(r_{0.002}, \,N_{0.002}, \,r_{0.05}, \,N_{0.05}, \,{n_s}_{0.05}) \simeq (0.0032, \,58.94, 0.0036, \,55.72, \,0.9649)$. Note that the value of the spectral index is precisely equal to the central value derived from the observations~\cite{Planck:2018jri}. Finally, as the parameter $\xi$ decreases, the predictions once again converge to the point ($\bullet$), remaining largely unaffected by the parameter $\widetilde{\xi}$.

In the lower panel, we set $\widetilde{\xi}$ to $10$ and vary $C$ from $0$ to $10^6$. In this case, the bullet ($\bullet$) indicates the case where $C=0$, consistent with the upper panel, while the stars ($\star$) represent the case where $C\gg \widetilde{\xi}^2$, corresponding to the results from~\cite{Ghilencea:2019rqj} as outlined by~\eqref{eq:KClimit}. In conclusion, a large value of $C$ can destroy the favorable inflationary predictions obtained when the $\widetilde{\xi}$ parameter is ``activated."

\section{Summary and Conclusions}

In the framework of Weyl-invariant Einstein-Cartan gravity we considered an action quadratic in the curvature scalars nonminimally coupled to a scalar field with a quartic potential. We analyzed the resulting system of scalar fields, namely, the above introduced fundamental scalar and the pseudoscalar field associated with the Holst invariant curvature term present in the action, focusing on the inflationary behavior of the system. We solved numerically the coupled system of the scalar equations of motion in an FRW background and found that for a range of the model parameters the system can evolve rapidly toward a trajectory in the two-field space that minimizes the potential. Along this line the potential depends only on the scalar field, while its kinetic energy receives a significant contribution from the pseudoscalar. Then, we proceeded to study the inflationary behavior of the resulting effective single-field system. The potential features a plateau in the small-field region with a minimum at larger field values. Although the potential coincides with that of ~\cite{Ghilencea:2019rqj}, there are significant deviations on  predictions due to the kinetic function, modified by the presence of the pseudoscalar. However, the possible solution of the strong \textit{CP} problem~\cite{Karananas:2024xja} associated with the latter corresponds to values of the parameters that do not fall into the range favored by the above evolution scenario.  Nevertheless, we do not rule out the possibility that the same theory could address both the \textit{CP} problem and inflation in regions of parameter space where the single-field analysis is not effective.

The fact that inflation occurs in the small-field region differentiates the model from its non-Weyl-invariant analog that realizes the {\textit{Higgs inflation}} scenario, having an inflationary plateau at large field values, despite the fact that the scalar field enters into the action as a nonminimally coupled Higgs-like scalar.  The inflationary predictions of the model occupy regions of parameter space where the nonminimal coupling between the Holst invariant and the scalar field, $\widetilde{\xi}$, is relatively large. It is important to note that large values of the standard nonminimal coupling $\xi$ are generally disfavored, as they reduce the available plateau region. Regardless of the value of $\xi$, increasing $\widetilde{\xi}$ drives the predictions for the spectral index and the tensor-to-scalar ratio toward a fixed point in the $n_s - r$ plane that lies well within observational bounds. Conversely, introducing a coupling between the axial (torsion) vector and the scalar field shifts these predictions away from the observed values.
\label{sec:conclusions}

%-------------------------------------------------------------------------------
\acknowledgments
%-------------------------------------------------------------------------------

The work of IDG was supported by the Estonian Research Council grants MOB3JD1202, RVTT3,  RVTT7, and by the CoE program TK202 ``Fundamental Universe''.

\appendix
\section{Mixing \texorpdfstring{$\cR\widetilde{\cR}$}{R tilde{R}}}
\label{appendix}
In the case that a mixing term $\frac{\varepsilon}{2}{\cal{R}}\widetilde{\cR}$ is present in ({\ref{SO}}) the equivalent action in terms of auxiliaries reads (for $\epsilon<\gamma\delta$, with analogous formulas in the opposite case)
\be
\frac{\gamma}{2}\chi{\cal{R}}+\frac{\delta}{2}\zeta\widetilde{\cR}
-\frac{1}{4}\left(\frac{\gamma\delta}{\gamma\delta-\varepsilon^2}\right)\left(\gamma\chi^2+\delta\zeta^2-2\epsilon\chi\zeta\right)\,.
\ee
Adding the matter sector~\eqref{DELTAS} and integrating out the non-dynamical degrees of freedom, as done previously, we find that the gravitational and kinetic parts of the action remain the same as in the unmixed case~\eqref{eq:S_2field}. However, the potential is modified to $V=V_0+\Delta V$, where  $V_0$ is given by~\eqref{eq:pot_V} under the substitutions $\g\rightarrow\g', \d\rightarrow\d'$, and 
\begin{align}
\Delta V=&-\frac{\epsilon\gamma'}{2\gamma^2}\zeta 
\\
=&\frac{-3\epsilon\gamma'}{\gamma^2\sqrt{\delta\delta'}(1+6\xi)}\left(\widetilde{\xi}-\frac{2e^{-(1+6\xi)\Phi}}{1-(1+6\xi)\tanh^2(H/\sqrt{6})}\right)\,, \nonumber
\end{align}
with $\gamma'=\gamma^2\delta/(\gamma\delta-\epsilon^2)$, $\delta'=\gamma\delta^2/(\gamma\delta-\epsilon^2)$ and  $\Phi= (1+6\xi)^{-1}\ln[(6+(1+6\xi)h^2)/(3\widetilde{\xi}-(\d\d')^{1/2}\zeta(1+6\xi)/2)]$.
Note that the extra term $\Delta V$ arising in the presence of ${\cal{R}}\widetilde{\cR}$ mixing vanishes along the minimum line~\eqref{eq:min_phi}.

\bibliographystyle{JHEP}
\bibliography{weyl_refs}{}
\end{document}